\documentclass[12pt]{article}
\input epsf.sty
\newcommand{\ra}{\rangle}

\newcommand{\ve}{\varepsilon}
\newcommand{\tl}{\wt\lambda}

\newcommand{\BB}{{\cal B}}

\newcommand{\HH}{{\cal H}}

\newcommand{\OO}{{\cal O}}

\newcommand{\LL}{{\cal L}}

\newcommand{\wt}{\widetilde}

\newcommand{\be}{\begin{equation}}
\newcommand{\ee}{\end{equation}}
\newcommand{\ben}{\begin{eqnarray}\displaystyle}
\newcommand{\een}{\end{eqnarray}}
\newcommand{\refb}[1]{(\ref{#1})}
\newcommand{\p}{\partial}
\newcommand{\sectiono}[1]{\section{#1}\setcounter{equation}{0}}

\begin{document}
{}~
\hfill\vbox{\hbox{hep-th/0203211}
\hbox{CGPG-02/3-2}
}\break

\vskip .6cm

\centerline{\Large \bf
Rolling Tachyon}

\medskip

\vspace*{4.0ex}

\centerline{\large \rm 
Ashoke Sen }

\vspace*{4.0ex}

\centerline{\large \it Harish-Chandra Research
Institute}

\centerline{\large \it  Chhatnag Road, Jhusi,
Allahabad 211019, INDIA}

\centerline {and}
\centerline{\large \it Department of Physics, Penn State University}

\centerline{\large \it University Park,
PA 16802, USA}

\centerline{E-mail: asen@thwgs.cern.ch, sen@mri.ernet.in}

\vspace*{5.0ex}

\centerline{\bf Abstract} \bigskip 

We discuss 
construction of classical time dependent solutions in open 
string (field) theory, describing the motion of the tachyon on unstable 
D-branes. 
Despite the fact that the string field theory action contains 
infinite number of time derivatives, and hence it is not a priori clear
how to set up the initial value problem, the theory contains a family of 
time dependent solutions characterized by the initial position and 
velocity
of the tachyon field. 
We write down the world-sheet action of the boundary conformal field 
theories associated with these
solutions and study the corresponding boundary 
states.
For D-branes in bosonic string theory, the energy momentum tensor of 
the system evolves asymptotically 
towards a finite limit
if we 
push the tachyon in the direction in which the potential has a local 
minimum, but hits a singularity if we push it in the direction where the 
potential is unbounded from below.

\vfill \eject

\baselineskip=16pt

\tableofcontents

\sectiono{Introduction} \label{s1}

Much work has been devoted to the study of tachyon potential and various 
classical solutions in string field theory on an unstable 
D-brane 
system\cite{sft,0005036}. However 
most of the 
solutions studied so far have been time independent solutions representing 
either the tachyon vacuum or static D-branes of lower dimension. Recently 
Gutperle and Strominger considered the process of production and decay of 
unstable branes, and proposed that the corresponding solution can be 
described as a space-like 
D-brane\cite{0202210}. For earlier attempt at studying the dynamics of 
tachyon on brane-antibrane system, see \cite{older}.

There are many issues associated with the decay of unstable branes in
string theory. We shall study the classical decay process, namely, what
happens if we displace the tachyon a little bit away from its maximum and
let it roll according to its equations of motion.
Although this is
a natural question in any scalar field theory around the 
maximum
of the potential,
it is not obvious if this is a sensible question in string 
field theory. The string field theory action contains infinite number of 
time derivatives, and hence {\it a priori} the initial value problem is 
not well-defined. Nevertheless, we shall show that it is possible to 
construct a family of classical solutions of the string field theory 
equations of motion characterized by the initial position and velocity of 
the tachyon field. The construction is similar in spirit 
to that of \cite{0202210} in that we get it by Wick rotation of a known 
solution, but the details are very different.

Another issue in the study of tachyon condensation is the nature of 
the time evolved configuration.
In studying the classical decay of an unstable soliton 
in a field 
theory ({\it e.g.} domain wall in $\phi^3$ theory\cite{0008227}), one can 
displace the 
field configuration from its original 
value and let it evolve in time. For simplicity we can restrict to 
deformations which do not depend on the coordinates along the brane, --
in terms of fields living 
on the brane this corresponds to constant field configuration. Typically 
in such a case the tachyonic field on the brane couples to various other 
fields, including those which are not localised on the brane and hence 
describes bulk modes. As a result even if the initial configuration 
contains only tachyonic deformation, with the energy density stored fully 
inside the tachyon field, as the system evolves in time the energy gets 
distributed among all the modes, including the infinite number of fields 
on the brane describing the bulk modes. In a generic case, if we wait 
long enough then 
eventually most of the energy is transfered to the bulk modes, and we can 
interprete this process as the result of the decay of the unstable brane 
into classical radiation.

In open string field theory, this issue is much more complicated. 
Unlike in an ordinary field theory, in this case we do not expect any 
conventional bulk modes, since there are no physical open string modes 
away from the original D-brane\cite{9911116,cohom,0007153}. Nevertheless 
one might 
expect 
that there 
are non-Fock space excitations of open string field configuration 
describing closed string 
excitations\cite{0111092,0111129,0011009,0203185}, and 
that the 
decay 
process 
transfers 
the energy of the original brane to these closed string excitations. 
We shall argue
that this is not what happens in classical open string field theory. 
Instead,
during the classical time evolution of 
the open string field describing the decay, the energy density
resides in 
the plane of the original brane. This situation is similar to that in the 
case of toy models of tachyon condensation\cite{toy} where again there are 
no 
bulk modes and the energy density does not dissipate\cite{private}.
We also find that the energy momentum tensor of the system evolves 
smoothly towards an asymptotic 
configuration instead of undergoing oscillation about the minimum of some 
effective potential. It is not clear how to reconcile this with the 
result from boundary string field theory that the minimum of the tachyon 
potential is at a finite distance away from the 
maximum\cite{0009103,0009148}.

One of the results that follows from the general analysis of this paper is 
that for D-branes in bosonic string theory, there is a qualitative 
difference between pushing the tachyon on the 
side of the maximum where there is a local minimum and the side where the 
potential is unbounded from below. In the boundary conformal field theory 
description, the former corresponds to adding a positive definite boundary 
term to the world-sheet action whereas the latter corresponds to adding a 
negative 
definite 
term. In the boundary state description, the former corresponds to a 
boundary state with smooth time evolution, while the latter hits a 
singularity at a finite time.

The rest of the paper is organised as follows. In section \ref{s2} we show 
the existence of a two parameter family of time dependent solutions, 
labelled by the initial position and velocity of the tachyon field, 
describing the rolling of the tachyon field on an unstable D-brane. In 
section \ref{s3} we 
construct the action of the boundary conformal field theory (BCFT) 
describing 
these solutions and in section \ref{sb} we explore the nature of the 
corresponding boundary states. In section \ref{s4} we study the decay 
process using 
an effective field theory. We conclude in section \ref{s5} with a few 
additional comments.

\sectiono{Time Dependent Solutions of Open String Field Equations} 
\label{s2}

In order to find a time dependent solution of the open string field 
equations, we can make a consistent truncation of the equations of motion 
restricting the fields to a universal subspace as follows. We shall assume 
that the original D-brane configuration is described by a unitary matter 
BCFT 
of central charge 25, together with a free scalar field 
representing the time coordinate $X^0$, and the ghost BCFT. Following the 
arguments of 
\cite{9911116,0005036} 
it is easy to show that if we restrict the open string field to the 
subspace of the full BCFT created from the vacuum by various operators 
associated with the scalar field $X^0$,
the Virasoro 
generators of the $c=25$ BCFT, and the ghost oscillators, then the 
equations of motion of the fields outside the subspace are automatically 
satisfied. Thus this provides a consistent truncation of the string field 
theory equations of motion, and we can look for the classical 
solutions 
describing time evolution of the brane within this subspace. Furthermore 
the action 
restricted to this subspace 
is universal, independent of the choice of the $c=25$ BCFT. This shows 
that the solution describing the decay of an unstable D-brane is 
universal, independent of which D-brane or spatial background we begin 
with.

Let us first consider the linearized equations of motion of string 
field theory where we ignore the coupling between various fields. In this 
case we can treat the tachyon as an ordinary scalar field of negative 
mass$^2$. 
For definiteness let 
us focus on the bosonic string theory, where in $\alpha'=1$ unit the 
tachyon has mass$^2=-1$. In this case, for small deformations the 
solution 
for the tachyon 
field will have the form 
\be \label{e1}
T(x^0) = A e^{x^0} + B e^{-x^0}\, ,
\ee
where $A$ and $B$ are constants determined by the initial condition:
\be \label{e2}
T(0) = \lambda, \qquad {\partial T\over \partial x^0}\Big|_{x^0=0} = 
u\, .
\ee
We shall for simplicity take $u=0$. 
Also, we shall take $\lambda>0$, so that $T$ rolls towards the extremum of 
the action
representing the tachyon vacuum, which is at a positive value of 
$T$\cite{sft}. This gives
\be \label{e3}
A=B={1\over 2} \, \lambda\, .
\ee
Hence
\be \label{e4}
T(x^0) = \lambda \cosh(x^0)\, .
\ee

As long as $\lambda$ is small and $x^0$ is not too large, we expect the 
linearized approximation to be valid, and the above 
represents the expected evolution of the tachyon field. The question now 
is: can we modify the above solution in a systematic manner to take into 
account the effect of the interaction? 
We can begin by  trying to construct a solution 
of the equations of motion:
\be \label{eeq}
Q_B |\Psi\ra + |\Psi*\Psi\ra = 0\, ,
\ee
as an expansion in 
$\lambda$:
\be \label{e6}
|\Psi\ra = \sum_n \lambda^n |\chi^{(n)}\ra\, .
\ee
The leading term is given by:
\be \label{e7}
|\chi^{(1)}\ra = c_1|0\ra_g \otimes \cosh(X^0(0))|0\ra_m\, .
\ee
The subscripts $g$ and $m$ refer to ghost and matter sectors respectively. 
We can construct the higher order terms in the solution in Siegel gauge
\be \label{e8}
b_0|\Psi\ra = 0\, ,
\ee
by choosing:
\be \label{e9}
|\chi^{(n)}\ra = -{b_0\over L_0} \sum_{m=1}^{n-1} |\chi^{(m)}*\chi^{(n-m)}
\ra\, .
\ee
This gives a way to construct the $n$th order term in terms of lower order 
terms.
However this procedure breaks down if $b_0\sum_{m=1}^{n-1} 
|\chi^{(m)}*\chi^{(n-m)}\ra$ contains a state with zero $L_0$ eigenvalue.
This is related to the question of whether the matter sector vertex 
operator
$\cosh(X^0)$ is exactly marginal. In this case, however, the 
marginality of this operator 
follows from the result in the Wick rotated theory where we replace 
$X^0$ by $i X$. $X$ is now a free scalar field describing a unitary 
conformal field theory of $c=1$. Under this rotation $\cosh(X^0)$ gets 
transformed to $\cos(X)$, and this is known to
represent an exactly marginal deformation in this Wick rotated 
theory\cite{marginal,9811237,9902105}. 
This shows that the original operator $\cosh(X^0)$ is also exactly 
marginal, and we do not encounter any obstruction in solving eq.\refb{e9} 
in a power series expansion in the parameter $\lambda$.
(If the coordinate $x$ were compactified on a circle of {\it critical
radius} $R=1$, the perturbation by $\cos(X)$ would create a single
codimension 1 D-brane on the circle.  However since the direction $x$ is
non-compact, this perturbation creates an array of codimension 1 D-branes
in the Wick rotated theory, seperated by an interval of $2\pi$.)

More explicitly we can proceed as follows. In the Wick rotated theory, the 
solution of the string field theory equations of motion 
for different values of the parameter $\lambda$
can be constructed 
explicitly 
using level 
truncation\cite{0007153}. 
We can simply take this solution and make an inverse Wick rotation $X\to 
-i X^0$ to 
generate a time dependent solution. One must of course 
make sure that Wick rotation of the solution preserves the reality 
condition, but this can be seen to be true as follows.\footnote{See 
\cite{9705038} for a detailed discussion on reality condition on string 
field.} The solution of 
ref.\cite{0007153} is invariant under the twist 
transformation (under which a string field component picks up a phase of 
$(-1)^l$ where $l$ denotes the level of the oscillators used to create the 
associated state from the zero momentum tachyon state $c_1|0\rangle$) as 
well as
the 
transformation $X\to -X$. Thus it can be expressed as linear 
combination of states, each of which only involves even total powers of 
the oscillator $\alpha_{-n}$ of $X$ and
momentum $k$ along the $X$ direction, and hence remains real under the 
inverse Wick 
rotation $k\to ik_0$, $\alpha\to -i \alpha^0$. Thus a real field 
configuration remains real after the inverse Wick rotation.

This shows the existence of a one parameter family of time dependent 
solutions, 
characterized by the marginal deformation parameter $\lambda$. A more 
general solution, labelled by two parameters, can be constructed by simply 
time-translating the original solution, and the two parameters can be 
interpreted as the initial position and velocity of the tachyon field. 
Thus we see that despite the non-locality of the string field theory 
action, it does admit time dependent solutions labelled by initial 
position and velocity of the tachyon field, as expected of an ordinary 
field theory 
with two derivative actions. 
However, since for all these solutions the tachyon comes to rest near the 
maximum of the potential at a 
certain instant of time, the total 
energy of each of these solutions is less than that of the unstable 
D-brane. 
Solutions for which the total energy is larger than that of the unstable 
D-brane can be constructed by beginning with $T(x^0)=u \sinh x^0$ 
and then iterating this using eq.\refb{e9}. 
The marginality of the operator $\sinh(X^0)$ again follows from the 
marginality of the operator $\sin(X)$ in the Wick rotated theory. Since 
under Wick rotation $u\sinh x^0$ becomes $iu\sin x$, and hence $T$ 
becomes imaginary, the time dependent solution is not related to any 
physical solution in the Wick rotated theory. 
But the marginality of the operator $\sin(X)$ 
\cite{marginal,9811237,9902105} would guarantee the existence of 
a complex solution in the Wick rotated theory whose leading term involves 
$T(x)=iu\sin(x)$. Under inverse Wick rotation this gives a real
solution in 
the original theory as a formal power series expansion in $u$.

Of course it is quite likely that all these solutions break down at 
sufficiently 
large $x^0$, since the higher order terms in the expansion proportional to 
$\cosh(n x^0)$ or $\sinh(n x^0)$ grow
with $x^0$. What we can hope for is that the solution exists at least for 
a finite range of values of $x^0$ during which we can use it to follow the 
time evolution of the string field. 
Since the expansion of the solution is in powers of $\lambda \cosh x^0$ 
(or $u \sinh x^0$), a 
rough estimate of the region of convergence for small $\lambda$ is 
$x^0<\ln(1/|\lambda|)$ (or $x^0<\ln(1/|u|)$). Beyond this limit we need to 
use other methods for studying the system, as will be discussed in 
sections \ref{s3} and \ref{sb}.
In this context we also note that even 
for $x^0=0$, the solution of \cite{0007153} can be constructed only 
for a 
limited range of $\lambda$: $|\lambda| \le \bar\lambda$ with $\bar 
\lambda\simeq .45$. This amounts to a 
restriction on the initial 
displacement of the tachyon field from its equilibrium value. 

Before we leave this topic let us remark that the solution obtained this 
way for a generic $\lambda$ is not the inverse Wick rotated version of a 
codimension one lump solution representing a lower dimensional D-brane. 
The codimension one lump 
representing a lower dimensional D-brane 
appears for a specific value $\lambda_c$ of the deformation parameter 
$\lambda$.\footnote{The relation between $\lambda_c$ and $\bar\lambda$ has 
not beeen conclusively determined\cite{0007153}.}
We shall argue in section \ref{sb} that this critical value $\lambda_c$ 
corresponds to placing the tachyon at the minimum of its potential in the 
Wick rotated theory.
For other values of $\lambda$, -- notably for small $\lambda$, 
-- the 
resulting conformal field theory is 
solvable\cite{marginal,9811237,9902105}, but 
cannot be interpreted as 
a D-brane of one lower dimension. This in turn shows that the time 
dependent 
solution 
for a generic $\lambda$ should not be thought of as an inverse Wick 
rotated version 
of a lower dimensional D-brane. Furthermore, even for $\lambda=\lambda_c$, 
the lump solution should be thought to be located not at $x=0$, but at 
$x=\pi$ where $\lambda \cos(x)$ reaches its minimum. Thus the origin of 
the $x^0$ coordinate, where the tachyon begins rolling, does not map under 
Wick rotation to the location of the lump solution.

We could repeat this construction for describing the decay of an unstable 
brane or a brane anti-brane pair in type II string theory. The arguments 
follow a similar line with the tachyon now having a mass$^2$ of 
$-1/2$ 
in the $\alpha'=1$ unit, and hence represented by a solution of the form
\be \label{es1}
T(x^0) = \lambda \cosh (x^0/\sqrt 2)\, ,
\ee
to first order in $\lambda$. 
In the Wick rotated theory the corresponding vertex operator is exactly 
marginal\cite{9808141}, which guarantees the existence of a solution in 
string 
field 
theory. 
The 
complete solution whose first order term is given by \refb{es1} has the 
property that the $\p_0 T$ vanishes at $x^0=0$. This reflects the fact 
that at $x^0=0$, the tachyon field has zero velocity. On the other hand 
the Wick rotated version of the solution, for which $T(x)\sim \lambda 
\cos(x/\sqrt 2)$ for small $\lambda$, should be thought of as an anti-kink 
located at 
$x= {\pi/ \sqrt 2}$, and a kink located at $x={3\pi/ \sqrt 2}$, {\it i.e.} 
at the points on the circle where the 
tachyon field itself vanishes. In particular if we increase $\lambda$, 
then for an appropriate value $\lambda_c$ of $\lambda$, it is at these 
points where the 
lower 
dimensional D-branes are created. Thus we see again that Wick rotation 
$x^0\to i x$ does not 
map the 
origin of the $x^0$ coordinate, where the tachyon begins 
rolling, to 
the location of the soliton solution.

Finally we note that both in the case of bosonic string theory and the 
superstring theory, we could get other solutions by inverse Wick rotation 
of the solutions describing lump / kink solutions on a circle of radius 
other than the critical radius where the deformation is marginal. 
The interpretation of 
these solutions is not entirely clear to us; however since at the critical 
radius they seem to represent the tachyon placed at its minimum, this may 
be the case for other radii as well.

\sectiono{Conformal Field Theory Description} \label{s3}

As pointed out already, the procedure outlined in the last section gives 
us a way to 
construct time dependent solution of the equations of motion for 
sufficiently small 
$x^0$, describing 
the initial stage of the brane decay, but does not really 
allow us to follow the time evolution for large $x^0$.
Thus we need to invoke other 
methods to analyze the system for large $x^0$.

One such method would be the techniques of two dimensional conformal field 
theory. As discussed already, the Wick rotated 
solution can be 
described by a solvable boundary conformal field 
theory\cite{marginal,9811237,9902105}, and 
hence one might 
try to describe the physics of the solution describing the rolling of the 
tachyon by analysing the inverse Wick rotation of this solvable conformal 
field theory. 
This is described by the world-sheet action:
\be \label{eff1}
-{1\over 2\pi} \, \int d^2 z \p_z X^0 \p_{\bar z} X^0 + \tl
\int dt 
\cosh X^0(t)\, ,
\ee
where $t$ parametrizes the boundary of the world-sheet, and  the 
deformation parameter $\tl$ of conformal field theory is related to the 
parameter $\lambda$ appearing in string field theory solution via some 
functional relation. For small $\lambda$, 
$\tl\simeq \lambda$\cite{asold,0007153}. We note from 
eq.\refb{eff1} that 
for small positive $\lambda$ (and hence positive $\tl$) the boundary term 
gives a positive 
contribution to 
the action. In contrast if we had chosen $\lambda$ to be negative, the 
boundary term will be negative, and could destabilize the theory. 
According to eq.\refb{e2} this corresponds to pushing the tachyon to the 
wrong side, where it can roll down all the way to $-\infty$. Presumably 
this is the instability that is seen in the conformal field theory 
description.\footnote{This qualitative difference between positive and 
negative values of $\tl$ is not visible in the Wick rotated 
theory\cite{0003101}.} For solutions with total energy larger than the 
brane tension, the $\cosh(X^0(t))$ term in \refb{eff1} is replaced by 
$\sinh(X^0(t))$ and we see an instability for either sign of $\tl$ 
since $\sinh(X^0(t))$ is unbounded from above and below. This is 
presumably related to the fact that in this case either in the past or in 
the future the tachyon rolls to the wrong side of the maximum where the 
potential is unbounded from below.

In contrast, for an unstable D-brane in superstring 
theory the solution describing the rolling of the tachyon is described by 
a boundary perturbation of the form\cite{9808141}:
\be \label{esuper}
\tl \, 
\int dt \,
\sigma_1\, \otimes \,
\psi^0(t) \sinh X^0(t)\, ,
\ee
where $\sigma_1$ is the Chan-Paton factor and $\psi^0$ is the superpartner 
of $X^0$. Thus both signs of $\tl$ are allowed in this case.

We shall not analyze these theories here, but note that since the Wick 
rotated version of these theories are 
solvable\cite{marginal,9811237,9902105,9808141}, these theories are likely 
to be 
solvable as well. We also note the following point.
Since the $\tl$-deformation of the original conformal field theory 
involves only the coordinate $X^0$, the part of the boundary conformal 
field theory involving the space-like coordinates remains unchanged under 
this deformation. In particular the 
coordinates transverse to the original brane, carrying Dirichlet boundary 
condition, will continue to carry Dirichlet boundary condition. Thus we 
would expect that as the configuration evolves in time, the energy density 
of the system is confined to the plane of the original brane. This can be 
seen in particular by probing the system by closed string vertex 
operators; if we take a closed string vertex operator carrying a factor of 
$e^{ik_\perp.X_\perp}$,
and compute its one point function on the disk with 
the boundary perturbation appropriate to the rolling solution, the answer 
is 
independent of $k_\perp$ due to the Dirichlet boundary condition on the 
coordinates $X_\perp$, or at most a polynomial in $k_\perp$ if the closed 
string vertex operator involves additional explicit factors of $k_\perp$ 
and/or $\partial X_\perp$. In 
position space this implies that the coupling of the closed strings
to the system is 
proportional to $\delta(x_\perp)$ or 
its derivative, indicating that the time dependent configuration is still 
firmly localized on the surface $x_\perp=0$.
This is in 
contrast to the generic decay of a soliton in a quantum field theory, 
where the initial energy of the lump soliton is eventually transfered to 
the bulk modes which carry the energy away from the brane.

\sectiono{Boundary State} \label{sb}

We shall now analyze the boundary state associated with the 
rolling tachyon solution to determine what kind of source it produces for 
closed strings. We shall restrict our analysis in this section to bosonic 
string theory.

We consider the rolling tachyon solution in open bosonic string theory 
associated with the 
world-sheet action \refb{eff1}. As usual we shall attempt to determine the 
associated boundary state by beginning with the boundary state in the Wick 
rotated theory and then making an inverse Wick rotation. The boundary 
state in the Wick rotated theory, described by a free non-compact scalar 
$X$ with boundary perturbation $\tl \int dt \cos(X(t))$ was analysed in 
\cite{9811237}. It is given by:
\be \label{eb1}
|\BB\ra \propto \sum_j \sum_{m=-j}^j D^j_{m,-m}(R) |j;m,m\ra\ra\, ,
\ee
where the sum over $j$ runs over $0, 1/2, 1, \ldots$, $R$ is the SU(2) 
rotation matrix:
\be \label{eb2}
R = \pmatrix{\cos(\pi\tl) & i \sin(\pi\tl) \cr i \sin(\pi\tl) & 
\cos(\pi\tl) }\, ,
\ee
$D^j_{m,-m}(R)$ is the spin $j$ representation matrix of this rotation in 
the $J_z$ eigenbasis, and $|j;m,m\ra\ra$ is the {\it Virasoro} Ishibashi 
state\cite{ishi} built over the primary $|j;m,m\ra$ in the $c=1$ CFT with 
momentum $2m$ and conformal weight $(j^2,j^2)$. This primary can be 
expressed in the form:
\be \label{eb3}
|j;m,m\ra = e^{i\ve(j,m)} \OO_{j,m} e^{2im X(0)}|0\ra_c\, ,
\ee
where $e^{i\ve(j,m)}$ is a phase factor, $\OO_{j,m}$ is a combination of 
oscillators of total level $j^2-m^2$ both in the holomorphic and the 
anti-holomorphic sector, and $|0\ra_c$ is the SL(2,C) invariant Fock 
vacuum for the closed 
string.

Since the expression is somewhat complicated, 
we shall focus on the part of the boundary state that does not involve any 
$X$ oscillators. This requires picking up only the primary states 
$|j;j,j\ra$ and $|j;-j,-j\ra$ in eq.\refb{eb1}, whose expressions, 
according to eq.\refb{eb3}, does not involve any oscillators since 
$j^2-m^2$ vanishes. 
Since
\be \label{eb3a}
D^j_{j,-j} = D^j_{-j,j} = (i\sin(\tl\pi))^{2j}\, ,
\ee
according to the convention of \cite{9811237}, 
the oscillator free part of the boundary state is given by:
\be \label{eb4}
|\BB_0\ra \propto \left[1+\sum_{j\ne 0} \, \, (i \sin(\tl\pi))^{2j} \, 
\left( 
e^{i\ve(j,j)} e^{2ij X(0)} + e^{i\ve(j,-j)} e^{-2ij X(0)} \right) 
\right] |0\ra_c\, .
\ee
It remains to determine the phase factors $e^{i\ve(j,\pm j)}$. These can 
be determined by using the result\cite{marginal,9811237,9902105} that at 
the special value of 
$\tl={1\over 2}$, the system becomes a periodic array of D-branes with 
Dirichlet boundary condition on $X$, placed at $x=(2n+1)\pi$. 
Alternatively, at $\tl=-{1\over 2}$ the system again becomes a periodic 
array of D-branes, but now placed at $x=2n\pi$. (In both cases the 
D-branes are
situated at the minimum of the potential $\tl\cos(X)$.) Thus for these 
values of 
$\tl$ the 
boundary state $|\BB\ra$ must match the 
known boundary state of the corresponding D-brane system. This gives
\be \label{eb5}
e^{i\ve(j,\pm j)} = (i)^{2j}\, ,
\ee
and hence, writing $j=n/2$, we get
\be \label{eb6}
|\BB_0\ra \propto \left[ 1+ 2\, \sum_{n=1}^\infty \, \, (- \sin(\tl\pi))^n 
\cos(n X(0))\right] |0\ra_c\, .
\ee
This shows that the source for the closed string tachyon (and other fields 
involving 
arbitrary excitations in the rest of the $c=25$ CFT and the ghost CFT but 
no oscillator 
excitation in the $c=1$ CFT) is proportional to
\be \label{eb7}
1+ 2 \, \sum_{n=1}^\infty \, \, (- \sin(\tl\pi))^n\,
\cos (nx)\, .
\ee
We can now make an inverse Wick rotation $x\to -i x^0$ to get the source 
for the closed 
string tachyon field (and other fields involving excitations in the $c=25$ 
and ghost
CFT) for the rolling open string tachyon solution. This is 
proportional to:
\be \label{eb7a}
f(x^0) \equiv 1+ \sum_{n=1}^\infty \, \, (- \sin(\tl\pi))^n \, (e^{nx^0} + 
e^{-n x^0})\, .
\ee 
This sum can be performed explicitly and gives:
\be \label{eb8}
f(x^0) = {1\over 1 + e^{x^0} \sin(\tl\pi)} + {1 \over 
1 + e^{-x^0} \sin(\tl\pi)} - 1\, .
\ee
Note that for positive $\tl$ (which corresponds to beginning on the side 
on which the potential has a local minimum) the function $f(x^0)$ is 
finite for 
all $x^0$ and in fact goes to 0 as $x^0\to\infty$.
On the other hand if $\tl$ is negative, {\it i.e.} we begin on the side 
where
the potential is unbounded from below, the function $f(x^0)$ hits 
a singularity at $x^0=\ln(-1/\sin(\tl\pi))$.

Another interesting quantity is the (Fourier transform of the) coefficient 
of the state $\alpha^0_{-1} \bar\alpha^0_{-1}|k^0\ra$ in $|\BB\ra$, where 
$\alpha^0_n$ and $\bar\alpha^0_n$ represent the 
oscillators associated 
with the right and the left-moving components of $X^0$. In 
the Wick rotated picture, this receives 
contribution from two sources: a constant term proportional to $D^1_{0,0}$ 
from the primary state $|1;0,0\ra$, and the secondary states ${1\over 2 
j^2} L_{-1}\bar L_{-1}|j;\pm j, \pm j\ra$ in $|j;\pm j, \pm j\ra\ra$. 
These contributions can be evaluated in a straightforward manner, and 
after inverse Wick rotation add up to a term proportional to
\be \label{ebf1}
g(x^0)=\cos(2\tl\pi) +1 - f(x^0)\, ,
\ee
with $f(x^0)$ defined as in eq.\refb{eb8}.
Like $f(x^0)$, $g(x^0)$ represents the $x^0$ dependence of the source for 
many closed string fields, involving excitation by 
$\alpha^0_{-1}\bar\alpha^0_{-1}$ in the $X^0$ CFT, and arbitrary 
excitations in the $c=25$ CFT and the ghost CFT. In particular the source 
for different 
components of the graviton 
and dilaton fields are given by appropriate linear combinations of 
$f(x^0)$ and $g(x^0)$. 

Since 
\be \label{ebf2}
f(x^0) + g(x^0) = \cos(2\tl\pi) + 1\, ,
\ee
is conserved, it is natural to interprete it as the conserved energy 
density on the D-brane up to an overall normalization.
This interpretation can also be supported by the following observation. 
For 
small $\tl$, we have
\be \label{ebf3}
f(x^0) + g(x^0) \simeq 2(1 - \tl^2 \pi^2)\, .
\ee
This should be compared with the initial total energy of the D-brane 
system when the tachyon field is displaced by a small amount 
$\lambda\simeq\tl$ from its maximum. The tension of the D-brane is given 
by $1/(2\pi^2 g^2)$\cite{9911116} where $g$ is the open string coupling 
constant. On the other hand, from the string field theory action we find 
that for small displacement $\tl$ of the tachyon field, the tachyon 
potential energy is $-\tl^2/(2 g^2)$. Thus the total initial energy is 
given by:
\be \label{ebf4}
{1\over 2\pi^2 g^2} ( 1 - \tl^2 \pi^2)\, .
\ee
This is precisely the same as the right hand side of \refb{ebf3} up to a 
constant of proportionality. This in turn shows that it is correct to 
interprete $f(x^0)+g(x^0)=1+\cos(2\tl\pi)$ as the total energy density of 
the D-brane 
system.

We end this section with the following observations:
\begin{enumerate}

\item For $\tl={1\over 2}$, total energy vanishes. Also from eq.\refb{eb8}
we see that $f(x^0)$ vanishes, indicating that the system does not evolve.  
This corresponds to placing the tachyon at the minimum of its potential.  
We also note that at this point the Wick rotated theory represents a
periodic array of D-branes with Dirichlet boundary condition on the $X$
coordinate, placed at $x=(2n+1)\pi$.

\item For $0<\tl<{1\over 2}$ the system evolves, but instead of 
oscillating about the minimum of the potential, its energy momentum tensor 
asymptotically 
approaches the configuration $f(x^0)\to 0$ and $g(x^0)\to 
(1+\cos(2\pi\tl))$. It is not clear how to reconcile this result with the 
result in boundary string field theory that the minimum of the tachyon 
potential  is a finite distance away from the 
maximum\cite{0009103,0009148}.

\item For $-{1\over 2}<\tl < 0$, the system evolves towards a singularity. 
As $\tl$ approaches $-{1\over 2}$, the time at which the system hits the 
singularity goes to zero.

\item The set of inequivalent solutions are obtained by taking $-{1\over 
2} < \tl \le {1\over 2}$. In particular $\tl={1\over 2}+\epsilon$ gives 
the same solution as $\tl={1\over 2}-\epsilon$.

\item We have so far analyzed only part of the 
boundary state. In order to get a complete picture we should also analyze 
the time evolution of the rest of the boundary state involving higher 
level states in the $X^0$ CFT.

\item The analysis can be easily repeated for perturbation by 
$\tl\sinh(X^0)$. In fact this is related to perturbation by 
$\tl\cosh(X^0)$ by a replacement $\tl\to -i\tl$, $X^0\to X^0 + i\pi/2$. 
Making these replacements in eqs.\refb{eb8}, \refb{ebf1} we get
\be \label{eb88}
f(x^0) = {1\over 1 + e^{x^0} \sinh(\tl\pi)} + {1 \over
1 - e^{-x^0} \sinh(\tl\pi)} - 1\, ,
\ee
and
\be \label{ebf11}
g(x^0)=\cosh(2\tl\pi) +1 - f(x^0)\, .
\ee
Thus the system hits a singularity at positive (negative) value of $x^0$ 
for {\it small} negative (positive) $\tl$. The total energy of the system 
is now proportional to $1+\cosh(2\tl\pi)$. These results are consistent 
with the interpretation that $\tl$ represents the time derivative of the 
tachyon field at the top of the potential.

\end{enumerate}

\sectiono{Effective Field Theory Analysis}  \label{s4}

In this section we shall use an effective field theory to explore the 
decay of an unstable brane.
The idea behind this analysis is 
as follows. As pointed out in  
refs.\cite{earlier,9901159,9909062,0002223,0005031,0009061,0010181,0010240}, 
the effective field 
theory 
describing the coupling of the tachyon to the Born-Infeld lagrangian 
admits electric flux tube solutions whose dynamics is identical to that of 
fundamental strings. Thus if we begin with a D-brane with a constant 
electric field switched on along one of its tangential directions, and 
then allow the tachyon field to roll down, then we 
would expect that the end product of the decay process will contain 
fundamental strings lying along the plane of the original brane, together 
with other decay products. If the other decay products reside in the 
degrees of freedom far away from the plane of the brane, we would expect 
that the hamiltonian describing the dynamics of the electric flux tubes 
will decouple from the hamiltonian describing the rest of the decay 
products. On the other hand if the other decay products reside in the 
plane of the brane, then there will not be any such decoupling. 

We shall 
now show that the decoupling of this type does not occur within the 
analysis based on an effective field theory, 
leading to the conclusion that the decay products reside in the plane of 
the brane. We shall begin with a general effective action on the 
world-volume of the unstable D-brane, involving a set 
of scalar fields $\vec T$, and a 
U(1) gauge field $A_\mu$. The scalar fields $\vec T$ include the tachyon, 
and could 
also include
other scalars like the transverse coordinates of the D-brane. We shall 
restrict to field configurations 
which are translationally invariant on the brane, work in the $A_0=0$ 
gauge, and allow only $A_1$ to be non-zero, thereby allowing a (time 
dependent) uniform electric field $E$ along the 1 direction given by
\be \label{e11}
E = \p_0 A_1\, .
\ee
We shall also make the simplifying assumption that the Lagrangian density 
depends only on the fields $\vec T$, $A_\mu$ and their first time 
derivatives.
{}From the general result of \cite{9908142,0005040,0005048} it follows 
that 
the theory in the 
presence of the electric field can be related to the theory in the absence 
of electric field by rescaling the 00 and the 11 components of the metric, 
changing the overall normalization of the action,
and replacing the ordinary products by Moyal product involving the 
coordinates $x^0$ and $x^1$. Since our field configurations are 
independent of $x^1$, the Moyal product is the same as the ordinary 
product, and we only need to take into account the rescaling of the 
metric and change in the overall normalization of the action. This gives 
the general form of the lagrangian density for space-independent field 
configurations to be:
\be \label{e12}
\LL = \sqrt{1-E^2} f({1\over \sqrt{1-E^2}}\, \p_0 \vec T, \vec T)\, ,
\ee
where $f(\p_0 \vec T, \vec T)$ denotes the lagrangian density in the 
absence of the electric field. If we define,
\be \label{e13}
f_i(\p_0 \vec T, \vec T) = {\partial f(\p_0 \vec T, \vec T) \over \partial 
(\p_0 T_i)}\, ,
\ee
then the momenta $\Pi$ conjugate to $A_1$ and the momenta $P_i$ conjugate 
to $T_i$ are given by, respectively,
\ben \label{e14}
\Pi &=& {\p \LL\over \p E} = -{E\over \sqrt{1-E^2}} \, f( {\p_0 \vec 
T\over
\sqrt{1-E^2}}, \vec T) + {E\over 1-E^2} 
\, 
\p_0 T_i \, f_i( {\p_0 \vec T\over
\sqrt{1-E^2}}, \vec T)\, \nonumber \\
P_i &=& {\partial \LL\over\p( \p_0 T_i)} = f_i( {\p_0 \vec T\over 
\sqrt{1-E^2}}, \vec T)\, .
\een
{}From this we get 
\be \label{e15}
\p_0 T_i = \sqrt{1-E^2} \, g_i(\vec P, \vec T)\, ,
\ee
where $\vec g$ is the inverse function of $\vec f$, {\it i.e.}
\be \label{e16}
g_i(\vec f(\p_0 \vec T, \vec T), \vec T) = \p_0 T_i\, .
\ee
Using eq.\refb{e15}, we can eliminate $\p_0 T_i$ in terms of $P_i$ in 
the first of eq.\refb{e14} and get
\be \label{e17}
\Pi
={E\over 
\sqrt{1-E^2}}\, h(\vec P, \vec T)\, ,
\ee
where
\be \label{e18}
 h(\vec P, \vec T) = \bigg(- f\Big(\vec g(\vec P, \vec T), \vec
T\Big) + \vec P \cdot \vec g(\vec P, \vec T)\bigg)\, .
\ee
This gives
\be \label{e19}
E = {\Pi \over \sqrt{\Pi^2 + (h(\vec P, \vec T))^2}}\, .
\ee

Let us now turn to the construction of the hamiltonian density. This is 
given by:
\be \label{e20}
\HH(\vec P, \vec T) = \Pi\,  E + \vec P\cdot \p_0 \vec T - \LL
= \sqrt{\Pi^2 + (h(\vec P, 
\vec T))^2}\, .
\ee
This shows that $h(\vec P, \vec T)$ represents the hamiltonian 
density of the system for $E=0$. $\HH$ is conserved as a result of time 
translation invariance of the system, and $\Pi$ is conserved as a result 
of gauge 
invariance. Thus from \refb{e20} we see that $h(\vec P, \vec T)$ is also 
conserved, and eq.\refb{e19} then shows that $E$ is conserved as well. We 
note from eq.\refb{e20} that due to separate conservation of 
$\Pi$ and $h$, the dynamics of $\vec P$ and 
$\vec T$ in the background of electric flux is related to the dynamics in 
the 
absence of the flux by a simple time dilation factor of 
$h/\sqrt{\Pi^2+h^2}$.

$\Pi$ can be interpreted as the energy stored in the electric flux tube; 
indeed if during the evolution of the D-brane the excess energy leaves the 
plane of the brane leaving behind static electric flux tubes in the 
tachyon vacuum, then the 
energy stored in the flux tube will be exactly 
$\Pi$\cite{0002223,0009061,0010240}. In this case 
we 
would expect the hamiltonian describing the dynamics of the flux tube to 
be nearly decoupled from the Hamiltonian describing the dynamics of the 
`bulk modes'. As seen from \refb{e20} however, this is not the case; the 
hamiltonian $h(\vec P, \vec T)$ containing the excess energy of the brane 
couples strongly to $\Pi$. To see this more  explicitly, we can compare 
two situations. First consider the situation where we have an initial 
condition $h(\vec P, \vec T)=0$. This represents a system with electric 
flux at the 
minimum of the tachyon potential. 
In the other case we begin with a system with the same amount of electric 
flux near the top of the potential 
well and let it roll down towards the minimum, so that the initial 
condition 
sets $h$ to a  fixed non-zero constant required to give the correct 
tension of the brane. 
If the excess energy of the brane is carried away from the brane during 
the decay leaving behind static electric flux tubes, then after a 
sufficiently long time, the dynamics of the flux tube in the two cases 
should be described effectively by the same hamiltonian.
To see if this is the case,
we now compare the response of these two systems to 
small fluctuation of the scalar fields $X^i$ describing transverse 
coordinates of the brane. This can be done by deforming $h$ 
from its given value 
by switching on small amount of momenta $P_i$ conjugate to $X^i$. 
Physically, around the tachyon vacuum solution, $P_i$ represents the 
momentum carried by the flux tube in directions transverse to the original 
brane\cite{0009061,0010240}.
Since $h(\vec P, \vec T)$ is conserved, irrespective of how long we wait,
the dependence of 
$\HH$ on $P_i$ will be different in the two cases, since we 
are Taylor expanding a function around two different points.
Thus the dynamics of the flux tube continues to depend on the initial 
energy of the system irrespective of how long we wait. {}From this we
conclude that the excess energy 
does not get carried away by radiation, but resides in the plane of the 
brane where it couples strongly to the electric flux tubes.

\sectiono{Conclusion}  \label{s5}

In this paper we have explored some aspects of time dependent solution in 
string field theory on an unstable D-brane describing the tachyon rolling 
away from the maximum 
of its potential. Due to the presence of higher derivative terms in the 
string field theory action it is not {\it a priori} clear how to construct 
such solutions. We have shown that it is indeed possible to construct a 
family of
solutions of the string field theory equations of motion labelled by the 
initial position and velocity of the tachyon field. Furthermore, one can 
also explicitly write down the boundary conformal field theory action and 
the boundary state 
associated with these solutions.

Unlike in the case of  decay of an unstable lump solution in a generic 
field theory, where under a small perturbation the energy of the lump gets 
converted to the classical radiation moving away from the original 
position of the lump, in the case of classical decay of a D-brane, the 
energy is stored in the plane of the original brane. This can be traced to 
the fact that there are no open string degrees of freedom away from the 
plane of the D-brane, and hence no classical mode of the string can 
carry away the energy.

The analysis of the boundary state shows that for D-branes in bosonic 
string theory, if we push the system 
towards the minimum of the tachyon potential, its energy momentum tensor 
evolves smoothly towards an asymptotic configuration. Instead, if we push 
the system in the other direction in which the potential is unbounded from 
below, the energy momentum tensor quickly hits a singularity.

Clearly there are many avenues left open for exploration. 
This includes a detailed study of the 
boundary states
describing the rolling 
of the tachyon field on unstable D-branes in
superstring theories. In the Wick rotated theory this was studied in 
ref.\cite{9903123}, but since it was expressed in terms of a different set 
of 
variables than the Wick rotated coordinate $X$, it is difficult to use it 
directly for the present analysis.
Another possible approach would be construction of 
the time dependent solutions in vacuum string field theory\cite{0012251} 
along the line
of \cite{vacsol}.
These studies may turn out to be useful in 
uncovering the role 
of tachyon condensation in cosmology.

\medskip

{\bf Acknowledgement}: I would like to thank B.~Zwiebach for useful 
discussions.
This work was supported in part by a grant 
from the Eberly College 
of Science of the Penn State University.

\end{document}